# Micro-Raman spectroscopy of graphene defects and tracing the oxidation process caused by UV exposure

Somayeh Gholipour[1], Maryam Bahreini[1]*, Mohamad Reza Jafarfard[1]
[1] School of Physics, Iran University of Science and Technology, Tehran, Iran
*Email: M_Bahreini@iust.ac.ir

**Abstract**

Raman spectroscopy is one of the widely-used methods in the analysis of various samples including carbon-based materials. This study aimed to identify the number of layers and defects in graphene using micro-Raman spectroscopy. More specifically, the study examined tracing the oxidation process of graphene under UV exposure. Investigation of the effect of the power density of the Raman excitation laser reveals a linear dependence between the ratio of $I_{2D}/I_G$ and the power density of the excitation laser. Also, the absence of peak D due to the increase in power density provides evidence for the non-destructive nature of micro-Raman spectroscopy. Given the value of $I_{2D}/I_G$, one of the parameters for determining the number of layers in graphene which has reached 1.39 at the edge, the findings indicate the possibility of an edge fold of single-layer graphene. During the oxidation process, the intensity and position of the D peak increase as a function of exposure time. Alterations in the graphene Raman spectrum, comprising the disappearance of the 2D peak and the appearance of the D peak, trace and confirm the oxidation process of the sample.

**Introduction**

Graphene is a nomenclature attributed to a flat single layer of carbon atoms tightly packed into a two-dimensional honeycomb lattice in which the overlap between the valence and the conduction bands happens in two points at the border of the Brillouin zone, named K and K' [1]. After the rise of graphene in 2004 [2], researchers have growingly directed their attention towards this material given its unique properties such as high electron mobility (150000 $cm^2$/V s) [3], superior thermal conductivity (up to 5300 W/mK) [4], high mechanical strength (up to 1 TPa) [5], and high optical conductivity [6]. The unique electronic and optical properties of graphene stem from its energy band structure with a linear scattering of electronic states near the Dirac points of the Brillouin zone. This stated graphene is referred to as a zero-gap semiconductor comprising electrons that behave like massless Dirac fermions [7]. Graphene-based structures with their unique optical and electric properties reflect an ideal platform for light-induced current generation and detection[8]. Likewise, graphene has brought about the development of devices such as graphene-based electrodes [9], lithium-ion batteries, and superconductors, because of its electronic properties and interactions with other atoms [10].

The extraordinary properties of graphene are highly contingent on its structural features such as the number of layers, the density of defects, doping [11], and the density of impurities. For instance, the presence of defects in the graphene lattice as a potential source of intervalley scattering can transform graphene from a metal to an insulator [12]. Alternatively, the mechanical attributes can be improved and the synthesis of new phases can be induced [13] by introducing controlled amounts of defects into the lattice. To intentionally modify the properties of graphene, chemical functionalization of graphene using its oxidation is considered a prevalent method. It should be noted that defects in graphene refer to anything that breaks the infinite symmetry of the carbon honeycomb lattice [14].

Graphene characterization and the production of devices based on it calls for various techniques such as scanning electron microscopy (SEM), transmission electron microscopy (TEM), and focused ion beam (FIB). Electron and ion beam irradiation in these methods can damage the sample. The radiation-induced defects may lead to significant deterioration of the electron and heat conduction properties [15].

Raman spectroscopy is an ideal tool for examining these systems given its fast and non-destructive nature, offering high resolution and providing chemical, structural, and electronic information [16]. This technique is one of the types of vibrational spectroscopy methods that are used in many fields such as cancer diagnosis [17, 18], classification of edible essential oils [19], characterization of carbon-based materials [16, 20], study on semiconductors[21, 22], etc. Vibrational optical spectroscopy is dependent on the specific wavelength of light absorption via vibrational levels of the molecule [23]. Raman scattering process is usually very weak [24] and some enhancement methods such as surface-enhanced Raman scattering (SERS) can dramatically improve its performance [25, 26]. Raman spectroscopy can provide informative data on the nature, type [14], and concentration of defects in graphene [1, 13] with the occurrence of defect-induced peaks.

Accordingly, this study investigates the number of layers and defects in graphene utilizing micro-Raman spectroscopy. Defects that appear in graphene are intentional or unintentional. The unintentional defects discussed here are edges and defects caused by synthesis. In particular, intentional defects are at the hub. The effect of the power density of the excitation laser as a possible factor in causing defects is examined. This study's concentration is further on examining the effect of UV light exposure on the oxidation of graphene. Such UV irradiation is a pervasively-used method of oxidizing graphene due to the generation of ozone. Following a novel approach, we trace its oxidation trajectory by the evolution of the Raman spectrum of graphene exposed to UV irradiation.

## Materials and Methods

Two samples of single-layer and bi-layer graphene were prepared on a silicon oxide substrate with a thickness of 300 nm, using the chemical vapor deposition (CVD) method. Micro-Raman spectroscopy of the samples was performed with a power density of 300 kW/cm$^2$ to 900 kW/cm$^2$ by a 532 nm laser beam, 60× microscope objective, and spot diameter of 12 μm using the Raman microscope (Technooran Company, Microspectrophotometer: ram-532-004). Bi-layer graphene was exposed to the radiation of a mercury lamp as a source of UV light with a power of 125 mW, and the evaluation in its Raman spectrum was investigated at different times of exposure.

## Results and Discussion

The spectrum of the inside regions of both samples was recorded with a power density of 700 kw/cm$^2$ and an irradiation time of 3s. In the spectrum of the single-layer graphene sample, the G peak at 1584 cm$^{-1}$ and 2D peak at 2684 cm$^{-1}$, and in the bi-layer graphene sample spectrum, the G peak at 1588 cm$^{-1}$ and 2D peak at 2708 cm$^{-1}$ were observed.

Figure 1 illustrates the Raman spectrum of the samples. Considering that the G peak is the stretching mode caused by carbon bonds, the intensity of this peak increases with the increase in the number of layers due to the participation of more carbon atoms[27]. The ratio of 2D peak to G peak was 3.4 in single-layer graphene and 0.92 in bi-layer graphene.

The examination of the width at half the maximum value of the samples indicated that the FWHM value of the 2D peak in the bi-layer sample is wider than that of the single-layer sample. Hao, Y., et al (2010) determined the width of the 2D peak proportional to the number of graphene layers [27]. In this study, the value of FWHM for the 2D peak in single-layer graphene was reported as 27.5±3.8 cm$^{-1}$ and in bi-layer graphene as 51.7±1.7 cm$^{-1}$. While the values obtained in our examined samples are 38 cm$^{-1}$ in single-layer graphene and 62 cm$^{-1}$ in bi-layer graphene.

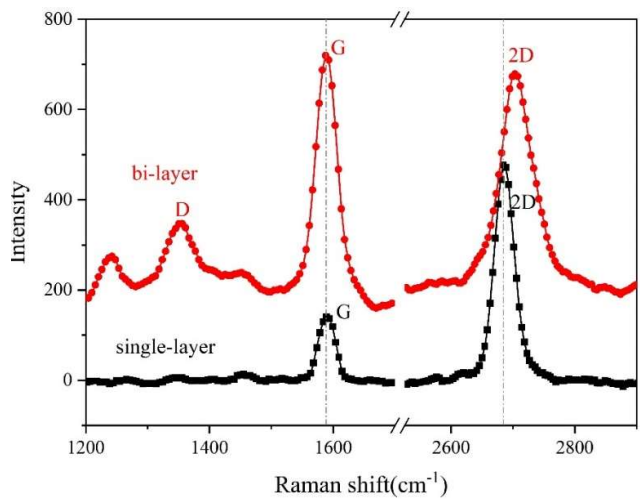

Fig. 1. Raman spectrum of single-layer and bi-layer graphene

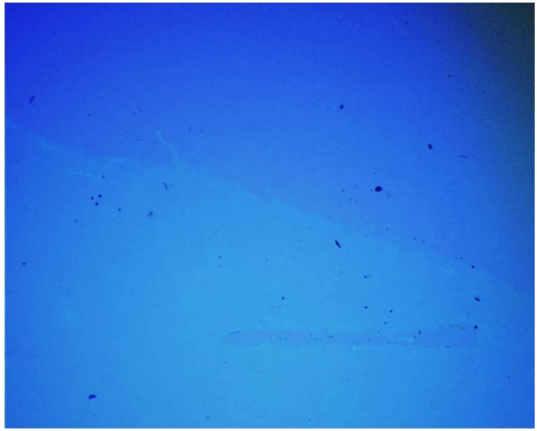

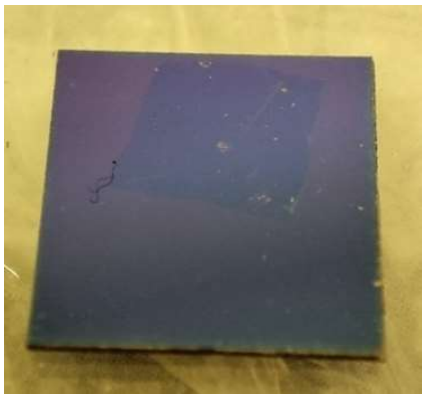

Fig. 2. (a) Optical microscope image of single-layer graphene sample. The contrast in the optical image confirms that graphene is a single layer. (b) Single-layer graphene.

*Edge; Unintentional defect*

The edge structure, notably in graphene nanoribbons, affects properties galore such as chemical reactivity, electronic structure, and vibrations. In structure identification, scanning tunneling microscopy can approach atomic resolution; however, edge structure analysis is often ambiguous. In this case, Raman spectroscopy is a valuable tool [28]. Figure 3 illustrates the Raman spectrum of the edge of the single-layer graphene sample compared to the Raman spectrum of its inside region. As presented in the figure, besides the main G and 2D peaks, the D peak appeared at 1350 cm$^{-1}$. The related literature on edges in graphene signifies the fact that a type of edge known as armchair edge whereby the arrangement of carbon atoms in the graphene structure is similar to a chair, shows a sharp and long Raman peak. It should be added that the spectrum related to the second type of edges known as zigzag edges shows a very weak D peak [29, 30]. Supporting the results reported in these studies, the edge investigated in our sample is of the armchair edge type. Another thing that was observed in the spectrum recorded from the edge is a significant increase in the FWHM and the height of the G peak compared to the G peak of the spectrum corresponding to the inside of the sample. The FWHM of the G peak increased from 33 inside the sample to 45.5 at the edge of the sample. $I_{2D}/I_G$ has also reached from the value of 3.4 inside the sample to the value of 1.39 at the edge. This ratio is close to the ratio of intensities in bi-layer graphene.

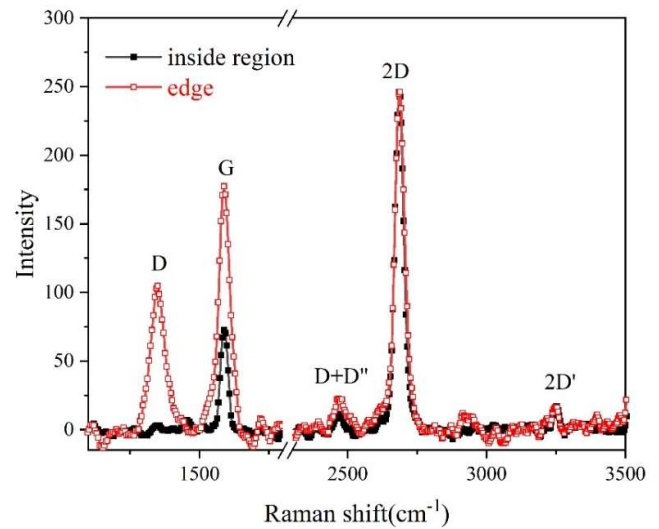

Fig. 3. Comparison of the Raman spectrum recorded from the inside region and the edge of the monolayer graphene sample

Another type of defect in graphene is the defect caused by the synthesis process. The D peak caused by this type of defect was observed in the Raman spectrum of the bilayer graphene sample at 1350 cm$^{-1}$ frequency (see Figure 1). The rise of this peak confirms the low quality of the synthesized sample. In the process of graphene synthesis by the CVD method, several reasons might be attributed to defects in graphene [31]. They are the oxidation of the sample due to the low reactivity of copper, the challenges of the transfer step from the copper substrate to the silicon substrate, and the entry of functional groups containing oxygen during the deposition into the carbon lattice.

*Intentional defect*

As noted earlier, analysis methods can also result in unintentional defects in graphene. The light source used in Raman spectroscopy is a laser. On the other hand, there are reports of the intentional destruction of graphene using low-power CO2 laser light [32], 248 nm pulsed laser [33], and 488 nm continuous laser [34]. The local heating created by the laser is a threat to the graphene under analysis. Since the emergence of some weak peaks is not possible at low powers of the laser light source, increasing the power becomes inevitable for better identification of the peaks. For this purpose, Raman spectra of single-layer graphene with different laser power densities were recorded and analysed. Figure 4 illustrates the result of the recorded spectra with laser exposure time 6s. From the start of spectroscopy with the lowest power density, i.e. 300 kw/cm$^2$, to the highest power density 900 kw/cm$^2$, the D peak caused by increasing the laser power was not observed. Therefore, it can be claimed that Raman spectroscopy up to 900 kw/cm$^2$ power density, which is considered a high power density for Raman measurements, does not damage graphene samples.

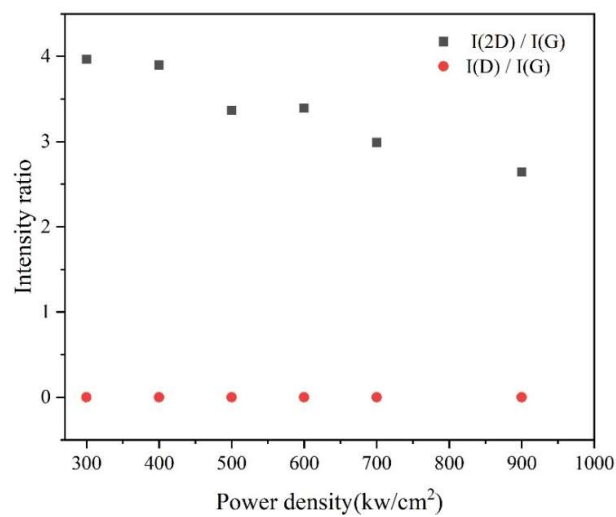

Fig. 5. Raman spectrum recorded from single-layer graphene with laser power densities of 300 to 900 kw/cm$^2$

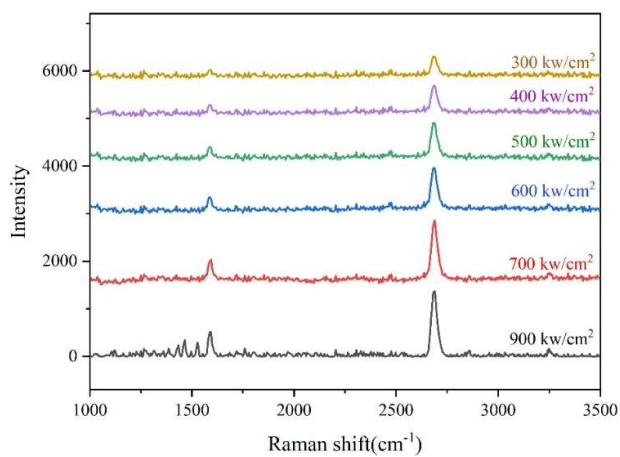

Fig. 4. Raman spectrum recorded from single-layer graphene with laser power densities of 300 to 900 kw/cm$^2$

On the other hand, despite the absence of peak D, a gradual increase in the intensity of peaks 2D and G was observed with increasing laser power density irradiation on the sample. The values of $I_D/I_G$ and $I_{2D}/I_G$ in each spectrum are shown in the diagram of Figure 5.

The $I_D/I_G$ values were always zero, while the $I_{2D}/I_G$ value went from 3.2 at 300 kw/cm$^2$ power density and 4 at 300 kw/cm$^2$ to 2.6 at 900 kw/cm$^2$ power density. As mentioned earlier, $I_{2D}/I_G$ determines the number of layers of the graphene sample. Despite the reduction of this value, it is still proportional to the single layer of graphene. With increasing laser power density, the G peak increases with a higher growth rate than the 2D peak. The reason for such behaviour can be attributed to the much higher sensitivity of the G peak to laser excitation energy. It should be noted that to ensure the accuracy of the results, this experiment was repeated three times at two different points. In each repetition, the laser power was altered from low to high and from high to low. According to the similarity of the results, the data related to one point is presented in the article.

Graphene oxide is considered a very defective type of graphene. This is a product of extensive oxidation of graphene, which does not have the electrical and thermal properties of graphene due to disrupting the structure of the graphene network. However, given the presence of oxygen groups, it shows a better interaction with different materials. One of the methods of graphene oxidation is the use of ozone, which is produced by UV light radiation in the environment. After producing ozone by UV light, ozone molecules react with graphene and produce O-containing groups, and induce p-type doping. Theoretically, O atoms lined up on pristine graphene are predicted to form a non-zippered structure where an epoxy group breaks the sp$^2$ bond of the bottom layer [35]. The effect of UV light produced by a 125 mW mercury lamp on the bi-layer graphene sample has also been reported in the following paragraphs. In Figure 6, the spectrum of the mercury lamp can be seen.

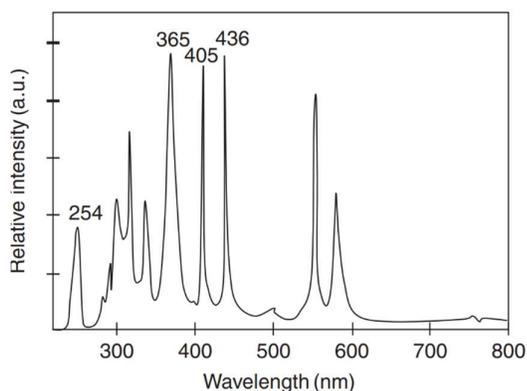

Fig. 6. Spectrum of mercury lamp

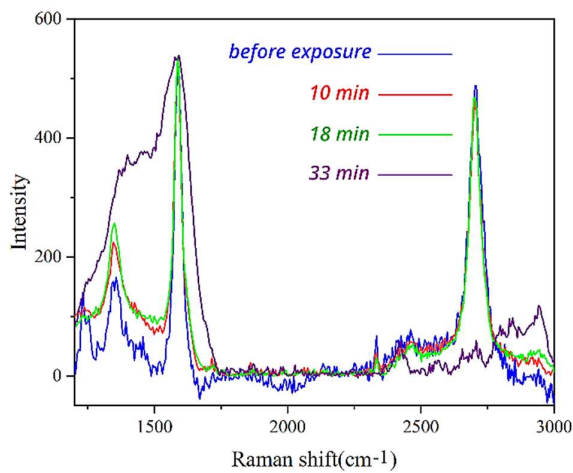

Fig. 7. Raman spectrum of graphene under UV light exposure at different times

In the oxidation of graphene, using UV light in the atmosphere, $O_2$ molecules absorb photons with a wavelength of 185 nm. The energy of these photons is sufficient to break molecular bonds and form two O(3P) in the ground state:

$$O_2 + h\nu = 2O(^3p) \text{ and } O(^3p) + O_2 = O_3$$

Furthermore, the produced $O_3$ can be destroyed in the reaction with O($^3$P): $O_3+O(^3P) = 2O_2$, so three species O($^3$P), $O_2$, and $O_3$ are involved in the oxidation process. O ($^3$P) can bond with carbon atoms and form an epoxy group [35]. The adsorbed epoxy groups create changes in the Raman spectrum of the sample by creating a shift in the Dirac cone, which ultimately leads to the oxidation of graphene.

Figure 7 depicts the evolution in the Raman spectrum of the sample before and after the start of exposure. During the UV exposure, the positions of the G and 2D peaks are roughly constant, but after 33 min, the G peak has a short red shift from 1588 cm$^{-1}$ to 1593 cm$^{-1}$ and the 2D peak has a long red shift from 2708 cm$^{-1}$ to 2941 cm$^{-1}$. As the exposure time increased, peak D appeared in the spectrum and its intensity increased over time. The appearance and growth of peak D indicate the gradual formation of epoxy groups in graphene.

The assessment of D peak intensity as a function of disorder in graphene is generally investigated by analyzing the behavior of the $I_D/I_G$ ratio. The intensity of the G peak did not change by disturbing the sample. Figure 5 shows the evolution of the intensity of peak D compared to peak G. With the oxidation of graphene, $I_D/I_G$ reached its maximum value. In addition, in the oxidized graphene, the G and D peaks were greatly broadened, so that after 33 minutes of exposure, these two peaks interfered with each other as presented in Figure 8.

Finally, the evolution of the two peaks G and D, the decrease in the intensity of the 2D peak, and the broadening of the G and D peaks confirmed the oxidation of the sample. In addition to the change in the intensity and broadening of peak D, the position of this peak also changed with the increase in exposure time and the formation of epoxy groups. Table 1 presents changes in the intensity and position of the D peak.

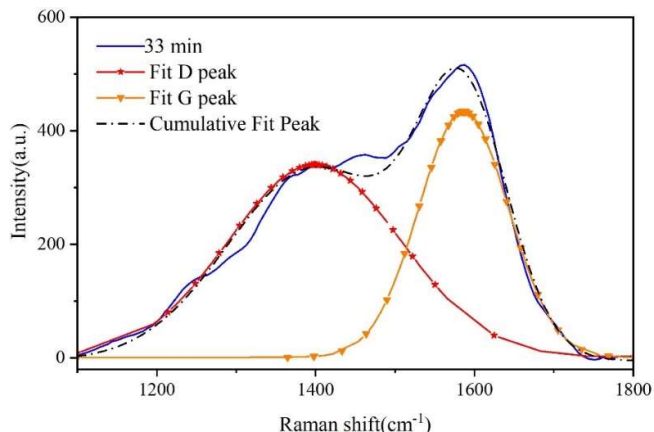

Fig. 8. Interference and broadening of G and D peaks in oxidized graphene

Table 1. Changes in the intensity and position of peak D due to the formation of epoxy groups in graphene with exposure to UV light

| $I_D/I_G$ | Position(cm$^{-1}$) | D peak |
|---|---|---|
| 0.22 | 1345 | 0 min |
| 0.25 | 1345 | 10 min |
| 0.36 | 1351 | 18 min |
| 0.52 | 1399 | 33 min |

## Conclusion

In this study, we detected the number of layers, edge defects, and defects caused by synthesis using micro-Raman spectroscopy. We also examined three parameters: the ratio of intensities, the position, and the FWHM of peaks. The results prove that the FWHM of the 2D peak parameter is not suitable for identifying the number of layers. Furthermore, there is a possibility of graphene folding at the edge of single-layer graphene due to the reduction of the $I_{2D}/I_G$. Likewise, we considered the effect of the power density of the excitation laser on the behavior of the Raman peaks of graphene and showed that $I_{2D}/I_G$ decreases linearly with increasing power density. Finally, we traced the Raman spectrum evolution of graphene under UV exposure. The alterations in the spectrum display the oxidation process appropriately. The appearance of the D peak indicates the initiation of the oxidation process. With the increase of exposure time and the growth of epoxy groups in the sample, the regular crystal structure of graphene was destroyed. Corresponding to the formation of epoxy groups, the D peak grew and eventually interfered with the G peak due to broadening. The 2D peak was reduced and had a long redshift to 2941 cm$^{-1}$.

## Author contribution statement

S. Gholipour and M. Bahreini carried out the experiments. S. Gholipour wrote the manuscript with support from M. Bahreini and M. R. Jafarfard. S. Gholipour and M. Bahreini fabricated the Raman Microspectrophotometer. M. Bahreini supervised the project. S. Gholipour and M. Bahreini conceived the original idea. S. Gholipour contributed to the interpretation of the results and analyzed the data. All authors provided critical feedback and helped shape the research, analysis and manuscript.

## Data Availability Statement

The datasets generated during and/or analyzed during the current study are available from the corresponding author on reasonable request.